\documentclass[11pt]{article}

\usepackage[margin=1in]{geometry}
\usepackage[T1]{fontenc}
\usepackage{lmodern}
\usepackage{amsmath,amssymb,amsthm,mathtools}
\usepackage{booktabs,microtype}
\usepackage{graphicx}
\usepackage[numbers,sort&compress]{natbib}
\usepackage[hidelinks]{hyperref}

\newtheorem{theorem}{Theorem}
\newtheorem{lemma}{Lemma}
\newtheorem{proposition}{Proposition}
\newtheorem{corollary}{Corollary}
\newtheorem{condition}{Condition}
\theoremstyle{remark}

\theoremstyle{definition}

\newcommand{\E}{\mathbb{E}}
\renewcommand{\Pr}{\mathbb{P}}
\newcommand{\G}{\mathcal{G}}
\newcommand{\IS}{\mathrm{IS}}
\newcommand{\ind}{\mathbf{1}}
\newcommand{\rbar}{\bar r}
\newcommand{\Dbar}{\bar\Delta}
\DeclareMathOperator{\Var}{Var}
\DeclareMathOperator{\KL}{KL}

\title{Comparing tail forecasts:\\the number of exceedances is the sample size}
\author{Raphael Coelho\\ Independent researcher, Brazil\\
\texttt{raphaelrrcoelho@gmail.com}\\ ORCID: 0009-0001-6601-1023}
\date{}

\begin{document}
\maketitle

\begin{abstract}
When two forecasts of a tail, such as prediction intervals at high
coverage, extreme quantiles or expected shortfall (ES), are compared by a
consistent score, the score difference is known in advance except on the
days on which the outcome falls beyond a forecast. The number of these
exceedances, not the number of observations, is the effective sample size.
For intervals at miscoverage level $\alpha$, the number of observations a
comparison needs grows like $1/\alpha$ while the number of exceedances it
needs does not, and for some pairs of forecasts no test can do better. A
confidence sequence that is valid at all times and needs no distributional
assumption attains this rate; Hoeffding-type procedures, calibrated by the
range of the score difference, need $1/\alpha$ times more. When
exceedances are scarce, the Diebold--Mariano test declares the less
conservative forecaster better with at least, asymptotically, the
probability of observing no exceedance, whichever forecaster is actually
better; the usual small-sample corrections do not remove this, while a
bootstrap-$t$ version largely does. For (value-at-risk, ES) forecasts, a
less conservative ES can be shown without assumptions to be worse on the
days on which it is less conservative, but never to be better without an
assumption on the loss tail. In 250-day windows of 97.5\% forecasts for
seven stock indices, 18.4\% of significant Diebold--Mariano results
favouring the less conservative model contradict the ranking on the
remaining days, against 6.9\% of those favouring the more conservative
one.
\end{abstract}


\section{Introduction}
\label{sec:intro}

Forecasts of rare events are issued as prediction intervals with high
nominal coverage, as quantiles at extreme levels, or as expected shortfall
(ES): 99\% intervals for electricity load or prices, the outer quantiles of
the weighted interval score used in epidemic forecasting
\citep{bracher2021evaluating}, and value-at-risk (VaR) and ES at the 97.5\%
level, on which market-risk capital rests \citep{bcbs2019minimum}.
Competing forecasters are compared through consistent scores: the interval
score \citep{winkler1972decision,gneiting2007strictly},
\begin{equation}
 \IS_\alpha(l,u;y)=(u-l)+\frac2\alpha(l-y)_++\frac2\alpha(y-u)_+,
 \label{eq:is}
\end{equation}
the quantile (pinball) score $\rho_\tau(y-q)$ with
$\rho_\tau(x)=x\{\tau-\ind(x<0)\}$, and, for ES jointly with VaR, the scores
of \citet{fissler2016higher} used in comparative backtests
\citep{nolde2017elicitability}. The usual tool is a test of equal
predictive ability on the score differences
\citep{diebold1995comparing,giacomini2006tests}, and practitioners know
that tail scores are noisy: tail-quantile scores ``suffer from high variance
due to the sparsity of observations'' \citep{browell2021probabilistic}.

This paper makes that observation precise. When the outcome falls inside
two intervals, their score difference equals the difference of their widths,
which is known before the outcome; it departs from that value only on
rounds in which an interval is missed (Lemma~\ref{lem:structure}). The same
holds for quantiles and for (VaR, ES) pairs with exceedances in place of
misses. Comparisons therefore rest on the exceedances. A 250-day
backtesting window at the 97.5\% level contains about six if the VaR
forecast is calibrated; a comparison of 99\% intervals over ten thousand
rounds rests on little more than a hundred. We call a procedure
\emph{anytime-valid} if its error guarantee holds simultaneously at all
sample sizes, so that the comparison can be monitored after every round and
stopped at any time.

We make three contributions.
\begin{enumerate}
\item \emph{How much data a comparison needs} (Sections~\ref{sec:cs}
and~\ref{sec:rates}). For some pairs of interval forecasts, every test
needs a number of misses that does not shrink with $\alpha$, and hence a
number of rounds that grows like $1/\alpha$, even for bounded outcomes. An
empirical-Bernstein confidence sequence, valid at all times with no
assumption on the outcomes, attains this rate up to constants and a
logarithmic factor. Procedures calibrated by the range of the score
difference need a factor of order $1/\alpha$ more.
\item \emph{Why the Diebold--Mariano test misleads} (Section~\ref{sec:dm}).
When misses are scarce it declares the narrower, or less conservative,
forecaster significantly better, whichever forecaster is better, with
probability at least, asymptotically, that of observing no miss; no test that
controls its error can reject this often.
\item \emph{What can be certified for ES} (Section~\ref{sec:es}). The score
difference is unbounded on one side whenever two ES forecasts differ. A
less conservative ES forecaster can be shown to be worse without
assumptions; no test can show it to be better without an assumption on the
loss tail.
\end{enumerate}
Simulations (Section~\ref{sec:sim}) confirm the rates and show the
Diebold--Mariano test, its usual small-sample corrections and an asymptotic
confidence sequence losing error control at small $\alpha$, while a
bootstrap-$t$ version of the test largely keeps it. Two applications follow: six online conformal
methods on seven forecasting data sets at levels from 80\% to 99\%
(Section~\ref{sec:empirical}), and six (VaR, ES) models on seven stock
indices (Section~\ref{sec:es-empirical}), where Diebold--Mariano tests in
250-day windows mostly favour the less conservative model and then often
contradict the ranking on the remaining days. Results stated for intervals
hold for quantile scores at tail level $\tau$ with $\alpha$ replaced by
$2\tau$, up to constants (Corollary~\ref{cor:interval-upper}); proofs
are in the Appendix.

\paragraph{Related work.}
Tests of equal predictive ability are reviewed by \citet{diebold2015comparing};
small-sample versions use the correction of
\citet{harvey1997testing} or fixed-smoothing asymptotics
\citep{coroneo2020comparing}, and the model confidence set
\citep{hansen2011model} extends the comparison to many forecasters. All
rely on a central limit theorem for the mean score difference, which
Section~\ref{sec:dm} shows to fail when exceedances are scarce. Rankings
by consistent scores can depend on the score chosen, which Murphy
diagrams display \citep{ehm2016quantiles}, and evaluation conditioned on
extreme outcomes is misleading \citep{lerch2017forecaster}; we fix the
score and condition on nothing. For ES,
\citet{fissler2016expected} and \citet{nolde2017elicitability} recommend
comparative backtests and note that rankings at extreme levels need long
samples.
\citet{bauer2025evaluating} found by simulation that Diebold--Mariano tests
of VaR and ES forecasts at the 1\% level have low power and wrong-sign
rejections favouring the less conservative forecast;
Proposition~\ref{prop:dm} gives a mechanism for the second finding.
\citet{frederiksen2026heavy} study a different failure, score differences
with infinite variance. \citet{jiao2025standard} construct comparative
e-backtests for risk measures including ES under a known bound on losses;
Proposition~\ref{prop:es-impossible} shows that some such assumption is
unavoidable in one direction. Confidence sequences for means go back to the
nonparametric constructions of \citet{howard2021time} and the betting
approach of \citet{waudbysmith2024estimating}. Anytime-valid comparison of
forecasters was developed by \citet{henzi2022valid} for binary outcomes and
by
\citet{choe2024comparing}, whose confidence sequences require bounded score
differences or a known global bound; they leave unbounded outcomes open and
suggest the asymptotic confidence sequences of \citet{waudbysmith2024time},
which we find lose error control at small $\alpha$.
\citet[Lemma~B.1]{arnold2026sequential} bound quantile-score differences by
a predictable multiple of the distance between the quantiles;
Lemma~\ref{lem:structure} is the interval-score version, with the
additional point that the deviation from a predictable centre vanishes off
the miss event. What is new here is the sample-complexity
analysis (lower bound, matching upper bound, the $1/\alpha$ separation
between variance- and range-calibrated procedures, and the
Diebold--Mariano mechanism) and the ES asymmetry.

\section{The structure of tail score differences}
\label{sec:setup}

Rounds are indexed by $t=1,2,\ldots$. Let $(\G_t)$ be a filtration in
which $\G_t$ is the information available when the forecasts for round $t$
are issued, and the outcome $y_t$ is $\G_{t+1}$-measurable. Two forecasters
$A$ and $B$ issue $\G_t$-measurable finite intervals
$I_{A,t}=[l_{A,t},u_{A,t}]$ and $I_{B,t}=[l_{B,t},u_{B,t}]$. Multi-step
forecasts fit this framework as long as the intervals for round $t$ are
issued before $y_t$ is observed. Lower scores are better, and we write
\[
 d_t=\IS_\alpha(I_{A,t};y_t)-\IS_\alpha(I_{B,t};y_t),\qquad
 \Delta_t=\E[d_t\mid\G_t],\qquad
 \Dbar_t=\frac1t\sum_{s=1}^t\Delta_s .
\]
The target $\Dbar_t$ is the average \emph{conditional} predictive ability
\citep{giacomini2006tests,choe2024comparing}: $\Dbar_t>0$ means that $B$ has
been better on average, round by round, given the information available
just before each outcome.
Let $H_t=\max\{|l_{A,t}-l_{B,t}|,|u_{A,t}-u_{B,t}|\}$ be the endpoint
distance and $p_{A,t}=\Pr(y_t\notin I_{A,t}\mid\G_t)$, and similarly
$p_{B,t}$, the conditional miscoverage probabilities. A \emph{miss} is a
round on which the outcome falls outside at least one of the two
intervals, $y_t\notin I_{A,t}\cap I_{B,t}$; when one interval contains the
other, the misses are those of the narrower one. For quantile and (VaR, ES)
forecasts we call the corresponding rounds \emph{exceedances}.

\begin{lemma}[Structure of interval-score differences]
\label{lem:structure}
Let $c_t=(u_{A,t}-l_{A,t})-(u_{B,t}-l_{B,t})$ be the width difference, and
\[
 m_t^-=\min\{0,\;l_{A,t}-l_{B,t},\;u_{B,t}-u_{A,t}\},\qquad
 m_t^+=\max\{0,\;l_{A,t}-l_{B,t},\;u_{B,t}-u_{A,t}\}.
\]
Then, for every value of $y_t$,
\begin{enumerate}
\item $d_t=c_t+(2/\alpha)f_t$, where $f_t$ is the distance from $y_t$ to
 $I_{A,t}$ minus its distance to $I_{B,t}$, and $m_t^-\le f_t\le m_t^+$, so that
 $a_t:=c_t+(2/\alpha)m_t^-\le d_t\le b_t:=c_t+(2/\alpha)m_t^+$;
\item $|f_t|\le H_t$, and $m_t^+-m_t^-\le 2H_t$;
\item $f_t=0$ if $y_t\in I_{A,t}\cap I_{B,t}$.
\end{enumerate}
The quantities $c_t$, $a_t$ and $b_t$ are $\G_t$-measurable. Consequently
$\Var(d_t\mid\G_t)\le\E[(d_t-c_t)^2\mid\G_t]
\le(4H_t^2/\alpha^2)\min\{1,p_{A,t}+p_{B,t}\}$.
\end{lemma}

When the intervals are nested, $I_{A,t}\subseteq I_{B,t}$, one has
$m_t^-=0$, so $d_t\ge c_t$: the score difference can only rise above the
width difference.

\paragraph{Quantile scores.}
For quantile forecasts $q_{A,t},q_{B,t}$ at a lower-tail level
$\tau<1/2$, $\rho_\tau(y-q)=\tau(y-q)+(q-y)_+$ gives
\begin{equation}
 d_t=\rho_\tau(y_t-q_{A,t})-\rho_\tau(y_t-q_{B,t})
 =\tau(q_{B,t}-q_{A,t})+\{(q_{A,t}-y_t)_+-(q_{B,t}-y_t)_+\}.
 \label{eq:quantile}
\end{equation}
The braced term lies between $\min\{0,q_{A,t}-q_{B,t}\}$ and
$\max\{0,q_{A,t}-q_{B,t}\}$ and vanishes unless
$y_t<\max\{q_{A,t},q_{B,t}\}$, an event of conditional probability at most
$\max\{p_{A,t},p_{B,t}\}$, where $p_{\cdot,t}=\Pr(y_t<q_{\cdot,t}\mid\G_t)$.
Upper-tail levels are symmetric. Since
$\IS_\alpha=(2/\alpha)\{\rho_{\alpha/2}(y-l)+\rho_{1-\alpha/2}(y-u)\}$,
interval scores are sums of two such comparisons.

\paragraph{A common condition.}
Both cases are instances of the following structure, which is all the
procedures below use.

\begin{condition}[Predictable centre and range]
\label{cond:range}
There are $\G_t$-measurable $a_t\le c_t\le b_t$ with $a_t\le d_t\le b_t$.
Write $D_t=d_t-c_t$, $R^-_t=c_t-a_t$ and $R^+_t=b_t-c_t$.
\end{condition}

For the rates we use two further quantities: a bound $\rbar$ with
$R_t^\pm\le\rbar$, and a bound $\pi$ with $\Pr(D_t\neq0\mid\G_t)\le\pi$.
Lemma~\ref{lem:structure} gives, for interval scores with $H_t\le H$ and
$p_{A,t},p_{B,t}\le C\alpha$,
\begin{equation}
 \rbar=2H/\alpha,\qquad \pi=2C\alpha,\qquad
 \E[D_t^2\mid\G_t]\le\rbar^2\pi=8CH^2/\alpha ,
 \label{eq:interval-constants}
\end{equation}
and, for quantile scores at level $\tau$ with
$|q_{A,t}-q_{B,t}|\le H$ and $p_{A,t},p_{B,t}\le C\tau$, $\rbar=H$ and
$\pi=C\tau$.

\section{An anytime-valid confidence sequence}
\label{sec:cs}

We now construct an empirical-Bernstein (EB) confidence sequence for
$\Dbar_t$. The idea is to measure the score difference against the predictable centre
$c_t$ and to estimate its variance only from the deviations $D_t$, which
are zero on every round without a miss. For a candidate value $m$ of the
average score difference and a tuning value $\lambda>0$, the quantity
$\exp\{\lambda\sum_s(d_s-m)-V\}$ is the wealth of a gambler who bets, with
stake $\lambda$, that the score differences exceed $m$, and $V$ is a
penalty for their fluctuations; if the true average were $m$, this wealth
could not grow large except with small probability. The lower bound $L_t$
is the largest $m$ that can be rejected in this way, using several stakes
at once and a one-dimensional root search. Formally, let
$g(u)=\{-\log(1-u)-u\}/u^2$ for $u\in(0,1)$, with $g(0)=1/2$ and
$g(u)=\infty$ for $u\ge1$; $g$ is increasing. Fix $\lambda_0>0$, set $\lambda_k=2^{-k}\lambda_0$ for
$k=0,\ldots,K-1$, and fix $\delta\in(0,1)$. For each $k$ and $m\in\mathbb R$
let
\[
 V_{k,t}=\lambda_k^2\sum_{s=1}^tg(\lambda_kR^-_s)D_s^2,\qquad
 \Phi_t(m)=\frac1K\sum_{k=0}^{K-1}
 \exp\Bigl\{\lambda_k\sum_{s=1}^t(d_s-m)-V_{k,t}\Bigr\},
\]
where a component is set to zero from the first round $s$ at which
$\lambda_kR^-_s\ge1$. The function
$\Phi_t$ is continuous and decreasing in $m$. Define the lower bound $L_t$ as
the root of $\Phi_t(m)=2/\delta$ (or $-\infty$ if every component is zero),
and the upper bound $U_t$ in the same way from $-d_t$, with the roles of
$R^-_t$ and $R^+_t$ exchanged.

\begin{theorem}[Validity]
\label{thm:valid}
Under Condition~\ref{cond:range} alone,
\[
 \Pr\bigl(\exists t\ge1:\;\Dbar_t\notin(L_t,U_t)\bigr)\le\delta .
\]
The same holds for any countable grid $(\lambda_k)$ with weights $w_k\ge0$,
$\sum_kw_k\le1$, in place of the uniform weights $1/K$.
\end{theorem}

No moment, tail, coverage or endpoint-distance assumption enters: the
ranges only need to be known before each outcome, and they are, because the
intervals are. Each evaluation of $\Phi_t$ costs $O(K)$ once the running sums are
updated, and the root is found by bisection. If $\lambda_0$, or the grid
and weights, are computed from data observed up to some round $t_0$, the
theorem applies conditionally on $\G_{t_0}$ to the rounds after $t_0$.

\paragraph{Choosing the grid.}
The uniform grid needs only the scale $\lambda_0$, which can be set from the
ranges $R_t^\mp$ in a short warm-up segment before evaluation; we call the
result the \emph{untuned} sequence, and Theorem~\ref{thm:upper} applies to
it. When the number of rounds $n$ to be evaluated is known, the weights can
instead concentrate on the $\lambda$ that suits it, about
$\{2\log(4/\delta)/(nv)\}^{1/2}$, where $v$, the mean of $D_t^2$, is
estimated on an initial tuning segment. The resulting \emph{tuned} sequence
is valid for the rounds after that segment (Appendix~\ref{app:tuned}),
which also lists the steps of its computation.
When misses are rare, the ranges say little about $v$, and the uniform grid
spends most of its weight on values of $\lambda$ far from the best one; the
tuned sequence avoids this at the cost of the tuning segment.

\section{How many rounds a comparison needs}
\label{sec:rates}

\subsection{Upper bound}

\begin{theorem}[Detection time]
\label{thm:upper}
Assume Condition~\ref{cond:range}, $R_t^\pm\le\rbar$ and
$\Pr(D_t\ne0\mid\G_t)\le\pi$ for all $t$. Let $\delta\le1/2$ and
$t\ge4$, and suppose $\lambda_0\ge1/(4\rbar)$ and
$2^{K-1}\ge\lambda_0\rbar\sqrt t$. Let
$\beta\in(0,1)$ and $\ell=\log(2K/\delta)+\log(2/\beta)$. If
$\Dbar_t\ge\Delta>0$ almost surely and
\begin{equation}
 t\ge\Bigl(96\,\frac{\rbar^2\pi}{\Delta^2}+22\,\frac{\rbar}{\Delta}\Bigr)\ell,
 \label{eq:upper}
\end{equation}
then $\Pr(L_t>0)\ge1-\beta$.
\end{theorem}

Substituting
\eqref{eq:interval-constants}:

\begin{corollary}[Interval scores]
\label{cor:interval-upper}
If $H_t\le H$ and $p_{A,t},p_{B,t}\le C\alpha$ for all $t$, the confidence
sequence separates the forecasters, $L_t>0$, with probability at least
$1-\beta$ as soon as
\[
 t\ge\frac{\ell}{\alpha}\Bigl(768\,C\,\frac{H^2}{\Delta^2}+44\,\frac H\Delta\Bigr).
\]
At that round the expected number of misses, at most $2C\alpha t$, is of
order $\ell C(CH^2/\Delta^2+H/\Delta)$, which does not depend on $\alpha$.
For quantile scores at level $\tau$ with $|q_{A,t}-q_{B,t}|\le H$ and
$p_{A,t},p_{B,t}\le C\tau$, the condition is
$t\ge\ell(96C\tau H^2/\Delta^2+22H/\Delta)$, with $\Delta$ in units of the
quantile score.
\end{corollary}

For interval scores $|\Delta_t|\le2H+(2/\alpha)H(p_{A,t}+p_{B,t})\le
(2+4C)H$, so $H/\Delta\le(2+4C)H^2/\Delta^2$, and the required number
of rounds is of order $(1+C)H^2\ell/(\alpha\Delta^2)$. The ratio $\Delta/H$ is a dimensionless effect
size: the average score difference relative to how far apart the forecasts
are.

\subsection{Lower bound}

\begin{theorem}[Lower bound for every test]
\label{thm:lower}
Let $\tau\in(0,1/2]$, $H>0$ and $0<\Delta\le\tau H$. There are fixed
quantile forecasts $q_A<q_B=q_A+H$ at level $1-\tau$ and two
distributions $P_0,P_1$ of an outcome with two support points such that, for
i.i.d.\ outcomes:
\begin{enumerate}
\item the expected quantile-score difference is $0$ under $P_0$ and $\Delta$
 under $P_1$;
\item under both, each quantile is exceeded with probability at most
 $2\tau$;
\item every test based on $y_1,\ldots,y_T$ whose errors under $P_0$ and
 $P_1$ are both at most $\delta<1/4$ requires
 \[
  T\ge\tau(1-\tau)\frac{H^2}{\Delta^2}\log\frac1{4\delta}.
 \]
\end{enumerate}
For interval scores at level $\alpha=2\tau$, taking
$I_A=[l,q_A]$ and $I_B=[l,q_B]$ with a common lower endpoint below both
support points gives a pair with endpoint distance $H$, miscoverage at most
$\alpha$, and expected interval-score differences $0$ and
$\Delta_{\IS}=\Delta/\tau$, and the bound reads
$T\ge2(1-\alpha/2)H^2\log\{1/(4\delta)\}/(\alpha\Delta_{\IS}^2)$.
\end{theorem}

The lower bound applies to anytime-valid procedures as well, since the
event that a level-$\delta$ confidence sequence excludes zero by round $T$
is a test. It matches Corollary~\ref{cor:interval-upper}: both are of order
$H^2\log(1/\delta)/(\alpha\Delta_{\IS}^2)$. In expected misses of the narrower
interval the bound is $(1-\tau)(H/\Delta_{\IS})^2\log\{1/(4\delta)\}$: no procedure
can separate two forecasters before observing a number of misses of order
$(H/\Delta_{\IS})^2\log(1/\delta)$, however many rounds that takes. The outcome
laws in the construction are bounded; rarity, not heavy tails, is the
obstacle.

\subsection{Range-calibrated procedures}

A Hoeffding-type confidence sequence uses the same predictable ranges but
not the variance: by Hoeffding's lemma,
$\exp\{\lambda\sum_s(d_s-\Delta_s)-\lambda^2\sum_s(b_s-a_s)^2/8\}$ is a
supermartingale for each $\lambda\ge0$, and a mixture over $\lambda$ gives
bounds $(L^{\rm H}_t,U^{\rm H}_t)$ in the same way as in
Section~\ref{sec:cs}.

\begin{proposition}
\label{prop:hoeffding}
If $b_s-a_s=r$ for all $s$, then for every grid and weights,
$L^{\rm H}_t\le\bar d_t-r\sqrt{\log(2/\delta)/(2t)}$, where $\bar d_t$ is
the average of $d_1,\ldots,d_t$. For nested intervals with
$H_t=H$, $r=2H/\alpha$, so the radius exceeds $2\Delta$ while
$t<H^2\log(2/\delta)/(2\alpha^2\Delta^2)$. If moreover the $d_t$ are
i.i.d.\ with mean $\Delta>0$ and variance $\sigma^2$, then for such $t$,
$\Pr(L^{\rm H}_t>0)\le\sigma^2/(t\Delta^2)$.
\end{proposition}

In the Gaussian example below, $\sigma^2/\Delta^2$ is of order $1/\alpha$,
so the bound on $\Pr(L^{\rm H}_t>0)$ is of order $\alpha$ at the
threshold, and the Hoeffding-type sequence needs of order
$1/\alpha$ times more rounds than the confidence sequence of
Section~\ref{sec:cs}.

The variance of $d_t$ is of
order $H^2/\alpha$ while its range is of order $H/\alpha$; the gap between
the two is the gap between the number of misses and the number of rounds.

\subsection{A Gaussian example}
\label{sec:gauss}

Let $y\sim N(0,1)$, let $z$ be its $(1-\alpha/2)$-quantile, and let $B$ be the exact
central interval $[-z,z]$. Forecaster $A$ either narrows or widens it by
$\eta/z$ at each end, with $\eta=1/2$. Then $H=\eta/z$, and
$\Delta$ and $\Var(d)$ follow from one-dimensional integrals of the normal
tail. As $\alpha\downarrow0$, $\Delta\asymp1/z$ and
$\Var(d)\asymp1/(\alpha z^2)$, so $\Delta/H$ tends to a positive constant
and $\Var(d)/\Delta^2\asymp1/\alpha$; for the widened interval, for
example, $\Delta\sim2(\eta-1+e^{-\eta})/z$ and
$\Var(d)\sim8\{1-(1+\eta)e^{-\eta}\}/(\alpha z^2)$. Section~\ref{sec:sim}
shows both rates in this example.

\section{The Diebold--Mariano test when misses are scarce}
\label{sec:dm}

The Diebold--Mariano statistic is
$\mathrm{DM}_T=\sqrt T\bar d_T/\hat s_T$, where $\hat s_T^2$ is a
(heteroskedasticity and autocorrelation consistent) estimate of the
long-run variance of $d_t$, and the test of equal predictive ability
rejects at level $\gamma$ when $|\mathrm{DM}_T|>z_{1-\gamma/2}$
\citep{diebold1995comparing,newey1987simple}. Its justification is a central
limit theorem for $\bar d_T$, which needs many misses.

\begin{proposition}
\label{prop:dm}
Let $(\sigma_t,\varepsilon_t)$ be i.i.d., with $\sigma_t>0$ independent of
$\varepsilon_t$, $\E\sigma_t^2<\infty$, and $y_t=\sigma_t\varepsilon_t$,
where $\sigma_t$ is known when the forecasts are issued. Let
$I_{A,t}=\sigma_t[-a,a]$ and $I_{B,t}=\sigma_t[-b,b]$ with $0<a<b$, both
scored by $\IS_\alpha$, and let $p=\Pr(|\varepsilon_t|>a)$. Let $\hat s_T^2$
be the sample variance or the Newey--West estimator with Bartlett weights,
autocovariances with divisor $T$ and lag $o(T)$, and declare $A$ significantly better when
$\mathrm{DM}_T<-z_{1-\gamma/2}$ (with $\mathrm{DM}_T=-\infty$ if
$\bar d_T<0=\hat s_T$). Then
\[
 \Pr\bigl(\mathrm{DM}_T<-z_{1-\gamma/2}\bigr)\ge(1-p)^T\,q_T,
\]
where $q_T$ depends only on the law of $(\sigma_t)$, not on $a$, $b$ or
the law of $\varepsilon_t$; $q_T\to1$ as $T\to\infty$, and $q_T=1$ if
$\sigma_t$ is constant. In particular, if $a=a_T$ varies with $T$ so that
$Tp_T\to\mu$, where $p_T=\Pr(|\varepsilon_t|>a_T)$, the probability is at
least $e^{-\mu}-o(1)$, for every value of $\Dbar_T$, including the null
value $0$ and values at which $A$ is worse.
\end{proposition}

With few misses, the score difference is dominated by the width
difference, which always favours the narrower interval, and the variance
estimate reflects only the variation in widths, so the statistic is large.
Where Theorem~\ref{thm:lower} says that no procedure can decide, the test
decides, in a direction fixed by the widths. The bound comes from the
event $N_T$ that no round up to $T$ is missed, and it identifies the
mechanism rather than its size: in the Gaussian design of
Section~\ref{sec:sim} with $\alpha=0.01$ and $T=250$, $(1-p)^T$ is below
0.1\%, while the test favours the narrower interval in 12\% of samples.
With a handful of misses the variance estimate is still dominated by the
width variation and by the few misses observed, and the distortion
persists as long as the expected number of misses is small.

\begin{proposition}[No valid test matches the Diebold--Mariano test]
\label{prop:dm-valid}
In the setting of Proposition~\ref{prop:dm} with $\sigma_t\equiv1$, let
$P$ be any law of $\varepsilon_t$ with $p=\Pr_P(|\varepsilon_t|>a)\ge\alpha$.
Let $\phi$ be any test based on $y_1,\ldots,y_T$ that declares $A$ better
with probability at most $\delta$ under every i.i.d.\ law of
$\varepsilon_t$ under which $A$ is not better. Then
\[
 \Pr_P(\phi\text{ declares }A\text{ better})\le\delta+1-(1-p)^T\le\delta+Tp,
\]
whereas the Diebold--Mariano test declares $A$ better with probability at
least $(1-p)^T$, under $P$ and under a law under which $A$ is not better.
\end{proposition}

When the expected number of misses $Tp$ is small, every valid test is
nearly powerless, while the Diebold--Mariano test rejects with probability
close to one whatever the truth: its additional rejections are errors, not
power. The condition $p\ge\alpha$ only asks that the narrower interval not
over-cover. The proposition applies to any procedure that controls its
error exactly, including the confidence sequence of Section~\ref{sec:cs};
procedures whose error control is only asymptotic, such as the
corrections studied in Section~\ref{sec:sim}, are not covered by it.

\section{Expected shortfall}
\label{sec:es}

Expected shortfall (ES) is not elicitable on its own, but the pair of
value-at-risk (VaR) and ES is \citep{fissler2016higher}, and (VaR, ES)
forecasts are compared through a strictly consistent scoring function for
the pair \citep{nolde2017elicitability}. We work with returns $y$, whose
lower tail is the loss tail, a level $\tau\in(0,1/2)$, and forecasts $(v,e)$
with $e\le v<0$ for the $\tau$-quantile and the $\tau$-expected shortfall,
scored by the FZ0 loss of \citet{patton2019dynamic},
\begin{equation}
 S(v,e;y)=-\frac{1}{\tau e}\ind(y\le v)(v-y)+\frac ve+\log(-e)-1 ,
 \label{eq:fz0}
\end{equation}
the member of the class of \citet{fissler2016higher} whose differences are
invariant to the scale of the returns. Write $\omega=1/(\tau|e|)$ for its
tail slope and $d_t=S(v_{A,t},e_{A,t};y_t)-S(v_{B,t},e_{B,t};y_t)$.

\begin{lemma}[Structure of FZ0 differences]
\label{lem:fz0}
Let $w(v,e)=v/e+\log(-e)$, $c_t=w(v_{A,t},e_{A,t})-w(v_{B,t},e_{B,t})$ and
$D_t=d_t-c_t$. Then
\begin{enumerate}
\item $D_t=\omega_{A,t}(v_{A,t}-y_t)_+-\omega_{B,t}(v_{B,t}-y_t)_+$, which is zero
 unless $y_t\le\max\{v_{A,t},v_{B,t}\}$;
\item if $e_{A,t}\ge e_{B,t}$, so that $A$'s ES is the less conservative,
 then $D_t\ge-\omega_{B,t}(v_{B,t}-v_{A,t})_+$, and if moreover
 $e_{A,t}>e_{B,t}$, then $D_t\to\infty$ as $y_t\to-\infty$; the case
 $e_{A,t}\le e_{B,t}$ is symmetric;
\item $w$ is strictly increasing in $|v|$ and nondecreasing in $|e|$, so
 $c_t\le0$ whenever $A$ is less conservative in both components,
 $v_{A,t}\ge v_{B,t}$ and $e_{A,t}\ge e_{B,t}$, and $c_t<0$ if moreover
 $v_{A,t}>v_{B,t}$.
\end{enumerate}
\end{lemma}

As for intervals and quantiles, the score difference departs from a value
known in advance only on exceedance days (item~1). Unlike them, it is
unbounded on one side whenever the ES forecasts differ (item~2). The same holds for every strictly consistent score in the
characterization of \citet{fissler2016higher}: the difference contains, for
outcomes below both VaRs, a term proportional to $\{G_2(e_A)-G_2(e_B)\}\,y$
with $G_2$ strictly increasing.

\paragraph{What can be certified.}
On days on which $A$'s ES is the less conservative, $d_t$ is bounded below,
so Theorem~\ref{thm:valid} applies to the lower bound; but large positive
deviations inflate its variance term. Flooring the outcome repairs this
without an assumption: replacing an extreme return by a less extreme one
can only lower $d_t$ on these days, so a lower bound computed from the
floored differences remains a valid lower bound for the original ones. Let $\underline y_t\le\min\{v_{A,t},v_{B,t}\}$ be
$\G_t$-measurable, and let $\tilde d_t$ be the score difference evaluated at
$\max\{y_t,\underline y_t\}$, which lies in a predictable bounded range.

\begin{proposition}[One-sided certification]
\label{prop:es-one-sided}
Let $\mathcal L=\{t:e_{A,t}\ge e_{B,t}\}$, a predictable set of days, and
let $\Dbar^{\mathcal L}_n$ be the average of $\Delta_t=\E[d_t\mid\G_t]$ over
the first $n$ days in $\mathcal L$. The lower bound $L_n$ of
Section~\ref{sec:cs}, computed from $\tilde d_t$ on the days in $\mathcal L$,
satisfies $\Pr(\exists n:\Dbar^{\mathcal L}_n<L_n)\le\delta/2$, with no
assumption on the distribution of the returns. The same holds for the upper
bound on the days in the complement of $\mathcal L$.
\end{proposition}

Thus one can show, without assumptions, that a forecaster whose ES is less
conservative was worse on the days on which it was less conservative. The
reverse cannot be shown without an assumption on the loss tail.

\begin{proposition}[No test can certify the less conservative ES]
\label{prop:es-impossible}
Fix forecasts $(v_A,e_A)$ and $(v_B,e_B)$ with $e_A>e_B$, used on every
day, and let the returns be i.i.d. Let $P_1$ be any law with a finite first
moment under which $\E_{P_1}d<0$, so that $A$ is better. Let $\phi$ be any
test based on $y_1,\ldots,y_T$ that rejects the hypothesis
$\E d\ge0$ with probability at most $\delta$ under every law with a finite
first moment for which that hypothesis holds. Then
$P_1(\phi\text{ rejects})\le\delta$.
\end{proposition}

The proof mixes $P_1$ with a small probability $\varepsilon$ of a very
large loss, which reverses the comparison while changing the law of the
sample by at most $T\varepsilon$ in total variation; this is the
obstruction of \citet{bahadur1956nonexistence}. Since the perturbation
changes the probability of an exceedance by at most $\varepsilon$, the
result holds even if the hypothesis is restricted to laws whose exceedance
probabilities for both VaR forecasts lie in any given open neighbourhood of
those under $P_1$, for instance laws under which both VaR forecasts are
nearly calibrated.
Certifying that a less conservative ES forecast is better therefore
requires an assumption on the loss tail. Under a known lower bound on the
returns, flooring the outcome at that bound changes nothing, the score
difference has a predictable range on all days, and the results of
Sections~\ref{sec:cs} and~\ref{sec:rates} apply.

\paragraph{Rates and the Diebold--Mariano test.}
Both carry over. With equal ES forecasts $e_A=e_B=e$ and $v_B=v_A-H$, the
FZ0 difference is $\omega\{-\tau H+\min((v_A-y)_+,H)\}$, the construction of
Theorem~\ref{thm:lower} scaled by $\omega=1/(\tau|e|)$, so separating the forecasters
requires at least $(1-\tau)\{H/(|e|\Delta)\}^2\log\{1/(4\delta)\}$
expected exceedances, where $H/|e|$ is the distance between the VaR
forecasts relative to the ES. On a sample without exceedances the score
difference equals $c_t$, which by item~3 of Lemma~\ref{lem:fz0} favours
the forecaster that is less conservative in both components, strictly when
its VaR is strictly less conservative. The proof of
Proposition~\ref{prop:dm} then applies with $c_t$ in place of the width
difference.

\section{Simulations}
\label{sec:sim}

\begin{figure}[t]
\centering
\includegraphics[width=\textwidth]{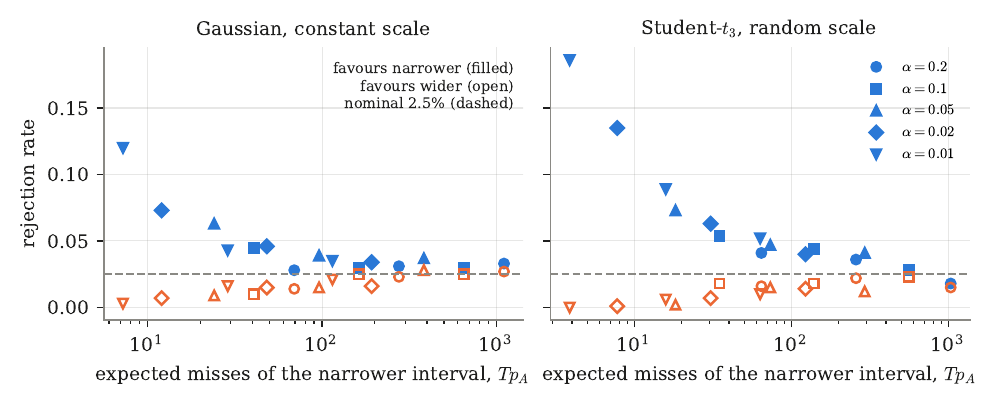}
\caption{Rejection rates of the two-sided 5\% Diebold--Mariano test under
equal expected scores, against the expected number of misses of the
narrower interval. Blue: the test declares the narrower forecaster better;
orange: the wider. Marker shapes give $\alpha$; the three points per shape
are $T=250$, $1000$ and $4000$.}
\label{fig:dm}
\end{figure}

\paragraph{Error control.}
In the first design the two forecasters have equal expected interval
scores, so $\Dbar_t=0$. Forecaster $A$ issues $[-k_A\sigma_t,k_A\sigma_t]$, where $k_A$ is
$0.85$ times the $(1-\alpha/2)$-quantile of the errors, and $B$ issues the
wider $[-k_B\sigma_t,k_B\sigma_t]$, with $k_B$ chosen so that both expected
scores coincide. Outcomes are standard Gaussian with $\sigma_t=1$, or
$y_t=\sigma_t\varepsilon_t$ with Student-$t_3$ errors and a lognormal scale
$\sigma_t$
(log-standard deviation $0.5$) known to both forecasters. For each
$\alpha\in\{0.2,0.1,0.05,0.02,0.01\}$ we run 1000 replications of length
4000 and evaluate the procedures at $T\in\{250,1000,4000\}$.
Table~\ref{tab:null} and Figure~\ref{fig:dm} report the results.

The empirical-Bernstein (EB) sequence never excluded zero in the
Student-$t$ design and did so in at most 0.1\% of replications in the
Gaussian design, and the Hoeffding sequence never did; the tuned
sequence of Section~\ref{sec:cs}, tuned on the first 10\% of each horizon
$T$ and evaluated at $T$, erred in at most 0.3\%. The asymptotic sequence of
\citet{waudbysmith2024time} is set to be tightest at $t=500$ and
monitored from round 31, since its variance estimate is degenerate in the
first rounds. It excluded zero at some round
in 2.5--4.8\% of Gaussian replications at $\alpha=0.2$, but in 58--60\% at
$\alpha=0.01$, and in 72--75\% at $\alpha=0.01$ in the Student-$t$
design. Most of its failures occur early in the sequence, while few
misses have occurred and the variance estimate is small; evaluated at the
final round only, its error is at most 2.2\% for $T\ge1000$, but up to
11.6\% at $T=250$.

The two-sided Diebold--Mariano test at 5\% has 2.5\% nominal error in each
direction. In favour of the narrower interval, its rejection rate at
$T=250$ is 12.0\% (Gaussian) and 18.6\% (Student-$t$) at $\alpha=0.01$,
and 7.3\% and 13.5\% at $\alpha=0.02$. In favour of the wider interval it
never exceeds 2.8\%. Figure~\ref{fig:dm} plots these rates against the
expected number of misses of the narrower interval: the points for
different $\alpha$ fall on one curve, and the excess over the nominal level
shrinks as that number grows. It exceeds twice the nominal level, 5\%,
in every case with fewer than 25 expected misses, and in the Student-$t$
design also at 31, 35 and 63.

\paragraph{Small-sample corrections.}
We also apply three standard modifications of the test, each two-sided at
5\%: the correction of \citet{harvey1997testing}, which rescales the
statistic and uses Student-$t$ critical values; the fixed-smoothing test
of \citet{coroneo2020comparing}, with a weighted periodogram variance
estimate on the $\lfloor T^{1/3}\rfloor$ lowest Fourier frequencies and
Student-$t$ critical values; and a bootstrap-$t$ test, which compares the
statistic with the quantiles of its distribution over 499 resamples of the
rounds, drawn independently as the outcomes are. Table~\ref{tab:dmcorr}
reports their rates in favour of the narrower interval. The first two
change the rates by at most 1.5 points, since they correct the variance
estimate, whereas the distortion comes from the skewness of $d_t$: most
rounds favour the narrower interval by a small amount and a few favour the
wider one by a large amount. The bootstrap-$t$ test captures this skewness
and keeps its rate at or below 3.6\% in every case except the smallest,
$T=250$ in the Student-$t$ design at $\alpha=0.01$, with four expected
misses, where it is 6.4\%; in favour of the wider interval its rate is at
most 3.7\% (not shown). It carries no finite-sample guarantee, and
Proposition~\ref{prop:dm-valid} does not apply to it.

\begin{table}[t]
\centering\small
\caption{Error rates (\%) under equal expected scores. DM: two-sided 5\%
Diebold--Mariano test at $T$, split by the forecaster it declares better.
Error by $T$: the 95\% confidence sequence excludes zero at some round up
to $T$ (asymptotic: from round 31). $Tp_A$: mean number of misses of the
narrower interval. 1000 replications.}
\label{tab:null}
\begin{tabular}{llrrrrrrr}
\toprule
 & & & \multicolumn{2}{c}{DM at $T$} & \multicolumn{3}{c}{error by $T$} & \\
\cmidrule(lr){4-5}\cmidrule(lr){6-8}
design & $\alpha$ & $T$ & narrower & wider & EB & Hoeffding & asympt. & $Tp_A$\\
\midrule
Gaussian & 0.2 & 250 & 2.8 & 1.4 & 0.0 & 0.0 & 2.5 & 69\\
 &  & 1000 & 3.1 & 2.3 & 0.0 & 0.0 & 4.0 & 276\\
 &  & 4000 & 3.3 & 2.7 & 0.0 & 0.0 & 4.8 & 1104\\
 & 0.1 & 250 & 4.5 & 1.0 & 0.0 & 0.0 & 8.9 & 41\\
 &  & 1000 & 3.0 & 2.5 & 0.1 & 0.0 & 10.4 & 162\\
 &  & 4000 & 3.0 & 2.5 & 0.1 & 0.0 & 11.5 & 649\\
 & 0.05 & 250 & 6.3 & 0.9 & 0.0 & 0.0 & 20.5 & 24\\
 &  & 1000 & 3.9 & 1.5 & 0.0 & 0.0 & 21.8 & 96\\
 &  & 4000 & 3.7 & 2.8 & 0.0 & 0.0 & 23.4 & 383\\
 & 0.02 & 250 & 7.3 & 0.7 & 0.0 & 0.0 & 41.2 & 12\\
 &  & 1000 & 4.6 & 1.5 & 0.1 & 0.0 & 42.6 & 48\\
 &  & 4000 & 3.4 & 1.6 & 0.1 & 0.0 & 43.6 & 192\\
 & 0.01 & 250 & 12.0 & 0.3 & 0.0 & 0.0 & 57.5 & 7\\
 &  & 1000 & 4.3 & 1.6 & 0.0 & 0.0 & 59.2 & 29\\
 &  & 4000 & 3.5 & 2.1 & 0.0 & 0.0 & 59.8 & 115\\
\midrule
$t_3$ & 0.2 & 250 & 4.1 & 1.6 & 0.0 & 0.0 & 4.5 & 64\\
 &  & 1000 & 3.6 & 2.2 & 0.0 & 0.0 & 6.6 & 258\\
 &  & 4000 & 1.8 & 1.5 & 0.0 & 0.0 & 7.1 & 1033\\
 & 0.1 & 250 & 5.4 & 1.8 & 0.0 & 0.0 & 12.5 & 35\\
 &  & 1000 & 4.4 & 1.8 & 0.0 & 0.0 & 13.8 & 139\\
 &  & 4000 & 2.8 & 2.3 & 0.0 & 0.0 & 15.3 & 557\\
 & 0.05 & 250 & 7.3 & 0.2 & 0.0 & 0.0 & 30.3 & 18\\
 &  & 1000 & 4.7 & 1.5 & 0.0 & 0.0 & 32.9 & 73\\
 &  & 4000 & 4.1 & 1.2 & 0.0 & 0.0 & 33.6 & 293\\
 & 0.02 & 250 & 13.5 & 0.1 & 0.0 & 0.0 & 51.6 & 8\\
 &  & 1000 & 6.3 & 0.7 & 0.0 & 0.0 & 54.3 & 31\\
 &  & 4000 & 4.0 & 1.4 & 0.0 & 0.0 & 55.0 & 123\\
 & 0.01 & 250 & 18.6 & 0.0 & 0.0 & 0.0 & 72.2 & 4\\
 &  & 1000 & 8.9 & 0.6 & 0.0 & 0.0 & 74.5 & 16\\
 &  & 4000 & 5.2 & 1.0 & 0.0 & 0.0 & 75.1 & 63\\
\bottomrule
\end{tabular}

\end{table}

\begin{table}[t]
\centering\small
\caption{Rejection rates (\%) in favour of the narrower interval under
equal expected scores, in the design of Table~\ref{tab:null}. DM: Diebold--Mariano
test with normal critical values; HLN: with the correction of
\citet{harvey1997testing}; WPE: fixed-smoothing test of
\citet{coroneo2020comparing}; bootstrap: bootstrap-$t$ test. All two-sided
at 5\%, so the nominal rate is 2.5\%. 1000 replications.}
\label{tab:dmcorr}
\begin{tabular}{llrrrrrr}
\toprule
design & $\alpha$ & $T$ & $Tp_A$ & DM & HLN & WPE & bootstrap\\
\midrule
Gaussian & 0.2 & 250 & 69 & 2.8 & 2.8 & 2.5 & 1.9\\
 &  & 1000 & 276 & 3.1 & 3.1 & 3.1 & 3.0\\
 &  & 4000 & 1104 & 3.3 & 3.3 & 3.3 & 3.6\\
 & 0.1 & 250 & 41 & 4.5 & 4.4 & 4.2 & 1.9\\
 &  & 1000 & 162 & 3.0 & 3.0 & 2.5 & 2.4\\
 &  & 4000 & 649 & 3.0 & 3.0 & 2.8 & 2.7\\
 & 0.05 & 250 & 24 & 6.3 & 6.1 & 4.9 & 0.9\\
 &  & 1000 & 96 & 3.9 & 3.9 & 3.9 & 2.5\\
 &  & 4000 & 383 & 3.7 & 3.7 & 3.4 & 2.9\\
 & 0.02 & 250 & 12 & 7.3 & 7.3 & 7.4 & 1.0\\
 &  & 1000 & 48 & 4.6 & 4.5 & 4.4 & 1.7\\
 &  & 4000 & 192 & 3.4 & 3.4 & 3.2 & 2.8\\
 & 0.01 & 250 & 7 & 12.0 & 11.9 & 10.8 & 0.8\\
 &  & 1000 & 29 & 4.3 & 4.3 & 4.0 & 1.2\\
 &  & 4000 & 115 & 3.5 & 3.5 & 2.9 & 1.5\\
\midrule
$t_3$ & 0.2 & 250 & 64 & 4.1 & 4.1 & 3.3 & 2.6\\
 &  & 1000 & 258 & 3.6 & 3.6 & 3.9 & 3.0\\
 &  & 4000 & 1033 & 1.8 & 1.8 & 2.3 & 2.4\\
 & 0.1 & 250 & 35 & 5.4 & 5.4 & 4.7 & 2.1\\
 &  & 1000 & 139 & 4.4 & 4.4 & 4.4 & 3.2\\
 &  & 4000 & 557 & 2.8 & 2.8 & 2.4 & 2.6\\
 & 0.05 & 250 & 18 & 7.3 & 7.3 & 6.6 & 2.3\\
 &  & 1000 & 73 & 4.7 & 4.7 & 4.4 & 2.6\\
 &  & 4000 & 293 & 4.1 & 4.1 & 3.9 & 3.1\\
 & 0.02 & 250 & 8 & 13.5 & 13.3 & 12.0 & 3.0\\
 &  & 1000 & 31 & 6.3 & 6.3 & 5.8 & 2.9\\
 &  & 4000 & 123 & 4.0 & 4.0 & 3.8 & 2.5\\
 & 0.01 & 250 & 4 & 18.6 & 18.4 & 17.4 & 6.4\\
 &  & 1000 & 16 & 8.9 & 8.9 & 8.3 & 1.1\\
 &  & 4000 & 63 & 5.2 & 5.2 & 4.8 & 2.9\\
\bottomrule
\end{tabular}

\end{table}

\begin{figure}[t]
\centering
\includegraphics[width=\textwidth]{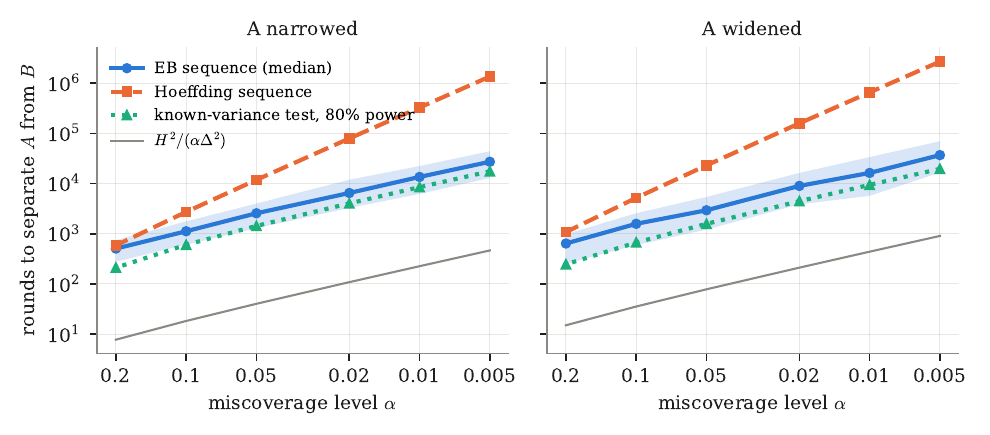}
\caption{Rounds needed to separate a narrowed or widened Gaussian interval
from the exact one. EB: median first round at which the 95\% lower bound
exceeds zero, with the 10\%--90\% band over 200 replications. Hoeffding:
round at which its deterministic radius falls below $\Delta$. Known-variance test:
sample size for 80\% power with known variance. Grey: the scale of the lower
bound, without constants.}
\label{fig:detection}
\end{figure}

\paragraph{Sample size.}
In the second design, outcomes are Gaussian, $B$ is the exact central
interval and $A$ is narrowed or widened by $\eta/z$ with $\eta=1/2$
(Section~\ref{sec:gauss}), so that $A$ is worse. For
$\alpha\in\{0.2,0.1,0.05,0.02,0.01,0.005\}$, 200 replications record the
first round at which the EB lower bound exceeds zero. Because the ranges are
constant, the Hoeffding radius is deterministic, and we report the round at
which it falls below $\Delta$. Detection times are recorded on a geometric grid of about 600
points.\footnote{The reported detection time is the first grid round at
which the lower bound exceeds zero, so it may exceed the exact first
round by at most the spacing of the grid, about 2\%.} Figure~\ref{fig:detection} also shows the
sample size at which a fixed-sample one-sided test at level 2.5\% with
known variance has 80\% power, $(z_{0.975}+z_{0.8})^2\Var(d)/\Delta^2$, and the scale
$H^2/(\alpha\Delta^2)$ of Theorem~\ref{thm:lower}, where $\Delta$ is the
expected interval-score difference ($\Delta_{\IS}$ in that theorem).

The EB and fixed-sample curves are parallel to $H^2/(\alpha\Delta^2)$: the
number of rounds grows like $1/\alpha$. The median number of misses at the
EB detection time is nearly constant, between 191 and 237 for the narrowed
interval and between 126 and 181 for the widened one, from $\alpha=0.2$ to
$\alpha=0.005$. The Hoeffding sequence grows like $1/\alpha^2$: it needs
1.2--1.7 times as many rounds as EB at $\alpha=0.2$ and 50--73 times as many
at $\alpha=0.005$, about $1.4$ to $2.7$ million rounds. Anytime validity
costs EB a factor between 1.6 and 2.6 relative to a fixed-sample test that
knows the variance and is valid only at the planned sample size.

With plentiful misses, as here, the Diebold--Mariano test is reliable: at
the sample size $T_{80}$ at which the known-variance test has 80\% power,
its power is 74--84\% with no wrong-sign rejection, and that of the
bootstrap-$t$ test 68--88\%, against 12--38\% for the tuned sequence
(Appendix~\ref{app:tuned}). Interpolating the power of the tuned sequence
between $T_{80}$, $2T_{80}$ and $4T_{80}$ on a logarithmic scale, it
reaches the power that the Diebold--Mariano test has at $T_{80}$ after
1.9 to 3.3 times as many rounds. This factor is close
to the ratio $\log(2/\delta)/(z_{0.975}^2/2)\approx1.9$
of the exponents of a Chernoff bound and of the normal approximation, which
is the price of a finite-sample guarantee.

\section{Application: interval forecasts at levels from 80\% to 99\%}
\label{sec:empirical}

\paragraph{Data.}
Online conformal methods provide many interval forecasters with the same
coverage target, which differ mainly in how they track the tails; they are
a natural test bed for comparisons at small $\alpha$. We use seven public
series: the oil temperature of an electricity
transformer, hourly (ETTh1) and every 15 minutes (ETTm1)
\citep{zhou2021informer}; the occupancy rate of one road sensor, hourly, and
one daily exchange rate \citep{lai2018modeling}; German--Luxembourg
day-ahead electricity prices, hourly, from Open Power System Data
\citep{wiese2019open}; and the first twenty series of the M4 hourly
\citep{makridakis2020m4} and NN5 daily \citep{godahewa2021monash}
collections, each concatenated into one series. After 50 warm-up rounds,
which initialise the methods and are not evaluated, the evaluation periods
have between 5{,}538 and 66{,}146 rounds; they are shorter at the 99\%
level, for the reason given below.

\paragraph{Forecasters.}
All methods share one point forecast and one scale. The point forecast
$\mu_t$ for $y_t$ is the mean of five simple forecasters (naive, seasonal
naive, rolling mean, simple exponential smoothing and a linear trend), each
using data up to round $t-h$, where $h$ is the seasonal period (24 for
hourly series, 96 for ETTm1, 7 for daily series). The scale $\sigma_t$ is
the standard deviation of the last 50 observations, floored at 5\% of the
series' standard deviation. Each method turns the standardized scores
$|y_s-\mu_s|/\sigma_s$, $s<t$, into an interval
$[\mu_t-\sigma_tq^-_t,\mu_t+\sigma_tq^+_t]$, so the intervals for round $t$
depend on information up to round $t-1$ only and satisfy the conditions of
Section~\ref{sec:setup}. The six methods are: adaptive conformal inference
aggregated over step sizes, AgACI \citep{zaffran2022adaptive}; dynamically
tuned adaptive conformal inference, DtACI \citep{gibbs2024conformal};
strongly adaptive online conformal prediction, SAOCP
\citep{bhatnagar2023improved}; quantile tracking (QT) and conformal PI
control (PI) \citep{angelopoulos2023conformal}; and AgACI-W, introduced
here, which aggregates adaptive conformal experts over several reference windows as
well as step sizes, with the ML-Poly rule of \citet{gaillard2014second}. For the methods
with a tuning parameter (the reference window of AgACI and DtACI, the
learning rate of QT and PI) we keep the configuration with the lowest mean
score on the evaluation period. The comparisons are therefore between the
best members of each family, selected in sample, and we use them to compare
the procedures rather than to rank the methods. At the 99\% level the
candidate reference windows are longer, and because they need more
history, the evaluation period of six series is 1{,}100 to 1{,}500 rounds
shorter at that level.

\paragraph{Procedures.}
For each data set and level we compare all 15 pairs of methods. The
Diebold--Mariano test is two-sided at 5\% with a Newey--West variance and
lag $\max\{h-1,\lfloor4(T/100)^{2/9}\rfloor\}$. The confidence sequences use
$\delta=0.05$. The EB sequence is the tuned one of Section~\ref{sec:cs},
tuned on the first 10\% of the evaluation period. The Hoeffding sequence
uses the uniform grid with $\lambda_0$ equal to the inverse of twice the
median range over the 50 warm-up rounds that precede the evaluation
period. We report whether each sequence excludes zero at
the last round and, for every procedure, which method it favours. As a check that uses
no model, we also split each evaluation period in halves, apply each
procedure to the first half (the EB sequence tuned on the first 10\% of
that half), and record whether the sign of its significant
results agrees with the mean score difference on the second half.

\paragraph{Results.}
Table~\ref{tab:empirical} summarizes the 105 comparisons at each level.
Misses, rounds on which at least one of the two intervals misses, make up
about $1.3\alpha$ of all rounds,
so at the 99\% level the median pair has 179 of them out of more than ten
thousand. The share of
pairs that each procedure separates falls with $\alpha$, fastest for the
range-calibrated Hoeffding sequence, which separates 44\% of pairs at 80\%
and none at 99\%, as Proposition~\ref{prop:hoeffding} predicts. The EB
sequence separates 70\% at 80\% and 18\% at 99\%, and the Diebold--Mariano
test declares 82\% and 54\% significant; the two never disagree in
direction. The untuned EB sequence, with $\lambda_0$ from the 50 warm-up
rounds as for the Hoeffding sequence, separates 59\% and 13\%.

The conditions of Section~\ref{sec:rates} roughly hold. Realized miss
rates have a median of $0.98\alpha$ to $1.00\alpha$ at every level (range
$0.6\alpha$ to $4.1\alpha$), and over windows of $20/\alpha$ rounds their
90th percentile has a median of about $1.1\alpha$ and is at most
$5.1\alpha$, so the conditional constant $C$ is small. Endpoint distances,
scaled by $\sigma_t$, have a 90th percentile about three times the median.
The effect size $\Delta/H$ has a median of 0.35, 0.37, 0.36 and 0.50 at
the four levels: it does not shrink with $\alpha$, so the $1/\alpha$ growth
of the required number of rounds applies to these data.

The split-sample check shows how reliable the significant results are. Of
the
Diebold--Mariano results that are significant on the first half, the share
whose sign agrees with the second-half mean falls from 100\% at 80\% to
97\%, 95\% and 90\% at 99\%. The sign of all 149 significant first-half
results of the EB sequence agrees, although at 99\% it has 15 such results
against 49 for the
Diebold--Mariano test. Configurations were selected on the whole period, so
the check is not fully out of sample.

\begin{table}[t]
\centering\small
\caption{Comparisons of six online conformal methods on seven data sets,
15 pairs each. Misses: rounds on which at least one interval misses,
median per pair. Significant: the
procedure separates the pair over the evaluation period (DM: two-sided
5\%; EB: tuned 95\% confidence sequence; Hoeffding: 95\% Hoeffding-type
sequence). First half:
number of significant results on the first half of the evaluation period,
and the share of those whose sign agrees with the mean on the second
half.}
\label{tab:empirical}
\begin{tabular}{rrrrrrrrr}
\toprule
 & & \multicolumn{3}{c}{significant (\%)} & \multicolumn{4}{c}{first half}\\
\cmidrule(lr){3-5}\cmidrule(lr){6-9}
 & & & & & \multicolumn{2}{c}{DM} & \multicolumn{2}{c}{EB}\\
\cmidrule(lr){6-7}\cmidrule(lr){8-9}
level & misses & DM & EB & Hoeffding & signif. & sign agrees (\%) & signif. & sign agrees (\%)\\
\midrule
80\% & 3905 & 81.9 & 69.5 & 43.8 & 72 & 100.0 & 58 & 100.0\\
90\% & 1982 & 66.7 & 47.6 & 26.7 & 66 & 97.0 & 40 & 100.0\\
95\% & 978 & 64.8 & 39.0 & 17.1 & 59 & 94.9 & 36 & 100.0\\
99\% & 179 & 54.3 & 18.1 & 0.0 & 49 & 89.8 & 15 & 100.0\\
\bottomrule
\end{tabular}

\end{table}

\section{Application: expected shortfall for seven stock indices}
\label{sec:es-empirical}

\paragraph{Data and forecasts.}
We use daily closing prices of seven equity indices (S\&P 500, FTSE 100,
DAX, Nikkei 225, Euro Stoxx 50, Hang Seng and Bovespa) from January 2000 to
September 2026, and percentage log returns. Six standard models produce
one-day-ahead (VaR, ES) forecasts at $\tau=0.025$ and $\tau=0.01$, each
using returns up to the previous day: historical simulation on the last 250
and 500 returns (HS-250, HS-500); the exponentially weighted variance with
decay factor 0.94 and normal quantiles (EWMA); GARCH(1,1) with normal (GARCH-N)
and standardized Student-$t$ innovations (GARCH-t); and filtered historical
simulation (FHS), which multiplies the GARCH-N volatility by the empirical
quantile and tail mean of the standardized residuals. The GARCH models have
zero mean and are re-estimated every 20 days on the last 1000 returns. After
an estimation window of 1000 days and a further 250 days that fix the
scale $\lambda_0$ of the untuned confidence sequence, the evaluation
periods have between 3{,}638 and 5{,}541 days.

\paragraph{Procedures.}
For each index and level we compare the 15 pairs of models by the FZ0 loss.
The Diebold--Mariano test is two-sided at 5\% with a Newey--West variance
and lag $\lfloor4(T/100)^{2/9}\rfloor$. We apply it to the full evaluation
period and to each non-overlapping window of 250 days, the length of the
regulatory backtesting sample. In each window we record whether a
significant result favours the model whose ES forecast was less
conservative on average over the window, and whether its sign agrees with
that of the mean loss difference over all other days of the evaluation
period. The certification of Proposition~\ref{prop:es-one-sided} is applied
on the days on which each model is the less conservative, with the outcome
floored at $\underline y_t=\kappa\min\{e_{A,t},e_{B,t}\}$ and $\kappa=2$,
that is, at twice the more conservative ES forecast; $\delta=0.05$, $K=20$,
and on the full period the sequence is tuned on the first 10\% of the
selected days. A tuning segment of 10\% of a 250-day window would contain
less than one expected exceedance, so within windows we use the untuned
sequence, with $\lambda_0$ from the 250 days that precede the evaluation
period. The choice of $\kappa$ affects power but not validity; we also
report $\kappa=1.5$ and $\kappa=3$.

\paragraph{Results.}
Table~\ref{tab:es} summarizes the results. A 250-day window contains a
median of nine days on which the return falls below at least one of the two
VaR forecasts at $\tau=0.025$, and four at $\tau=0.01$, against about six
and 2.5 for one calibrated forecast, because two forecasts are involved and
the models' exceedance rates, pooled over indices, are 1.0--1.4 times the
nominal level at $\tau=0.025$ and 1.1--2.1 times at $\tau=0.01$. The Diebold--Mariano test
is significant in 18\% of the window--pair tests at both levels, and 73\%
(at $\tau=0.025$) and 80\% (at $\tau=0.01$) of the significant results
favour the less conservative model. Over the full period, with a median of
203 and 104 exceedances per pair, the corresponding shares are 67\% and
51\%, counting a model as less conservative when its ES was the less
conservative on most days. The tilt towards the less conservative model is
thus modest at $\tau=0.025$, 73\% in windows against 67\% over the full
period, and large at $\tau=0.01$, 80\% against 51\%, where windows hold
the fewest exceedances. The significant window results that favour the less
conservative model contradict the ranking on the other days in 18.4\% and
21.1\% of cases; those that favour the more conservative model do so in
6.9\% and 10.5\%.

The window tests are dependent: the 15 pairs of an index share the returns,
and windows of different indices overlap in time. To assess the sampling
uncertainty of these shares, we assign each window to the calendar year of
its midpoint and resample years with replacement, keeping every window of
the drawn years (4000 resamples). The 95\% percentile interval for the
difference between the two wrong-sign rates is $[1.7,19.7]$ percentage
points at $\tau=0.025$ and $[0.0,18.9]$ at $\tau=0.01$; for the share of
significant results that favour the less conservative model it is
$[54,89]\%$ and $[67,90]\%$. The asymmetry is thus clear at
$\tau=0.025$ and borderline at $\tau=0.01$, where significant results of
either kind are fewer.

Replacing the normal critical values of the Diebold--Mariano test by those
of a circular block bootstrap-$t$ test (blocks of 10 days, 999 resamples,
with the statistic and its replicates studentized by the variance of the
block means) removes most of the tilt. The share of significant window
tests is similar, 18\% and 20\%, but only 51\% and 60\% of them favour the
less conservative model. Their wrong-sign rates are 12.9\% and 14.2\%,
against 9.5\% and 6.3\% for those that favour the more conservative
model, so at $\tau=0.01$ results
in favour of the less conservative model remain the less reliable.

The certification never succeeds within a window, as the lower bound of
Section~\ref{sec:es} leads one to expect on four to nine exceedances. On
the full period it certifies that the less conservative model was worse
on its own days for 48\% of pairs at $\tau=0.025$ and 44\% at
$\tau=0.01$. Across $\kappa\in\{1.5,2,3\}$ these shares range over 48--50\%
and 42--44\%. The untuned sequence certifies 43\% and 24\%. When it certifies, the
full-period
Diebold--Mariano test is significant in the same direction in 48 of 50 and
36 of 46 pairs, and never in the opposite direction.

\begin{table}[t]
\centering\small
\caption{Comparisons of six (VaR, ES) models on seven equity indices,
15 pairs each, by the FZ0 loss. Windows: non-overlapping 250-day windows.
Exceedances: days on which the return falls below the larger VaR forecast.
DM: two-sided 5\% Diebold--Mariano test with normal critical values. Wrong
sign: share of the significant window results, of each kind, whose sign is
opposite to that of the mean over all other days. EB certifies: share for
which the one-sided 95\% bound of Proposition~\ref{prop:es-one-sided}
certifies that the less conservative model was worse on its own days, with
the outcome floored at twice the more conservative ES; untuned sequence in
windows, tuned sequence over the full period.}
\label{tab:es}
\begin{tabular}{lcc}
\toprule
 & $\tau=0.025$ & $\tau=0.01$\\
\midrule
\multicolumn{3}{l}{\emph{250-day windows}}\\
window--pair tests & 2130 & 2130\\
median exceedances & 9 & 4\\
DM significant (\%) & 17.8 & 18.0\\
\quad favouring the less conservative (\%) & 73.1 & 80.2\\
\quad wrong sign, favouring less conservative (\%) & 18.4 & 21.1\\
\quad wrong sign, favouring more conservative (\%) & 6.9 & 10.5\\
EB certifies (\%) & 0.0 & 0.0\\
\midrule
\multicolumn{3}{l}{\emph{full period}}\\
pairs & 105 & 105\\
median exceedances & 203 & 104\\
DM significant (\%) & 77.1 & 71.4\\
\quad favouring the less conservative (\%) & 66.7 & 50.7\\
EB certifies (\%) & 47.6 & 43.8\\
\bottomrule
\end{tabular}

\end{table}

\section{Discussion}
\label{sec:discussion}

In tail forecast comparisons the effective sample size is the number of
exceedances. Three practical recommendations follow. First, report the
number of rounds on which the score difference departs from its predictable
value; it is computable from the forecasts and the outcomes. Second, do not
use normal critical values for the Diebold--Mariano test when this number
is small: in our simulations, its error in favour of the narrower interval
exceeded twice the nominal level in every design with fewer than 25
expected misses of the narrower interval, and with heavy-tailed outcomes in
some designs with up to 63. The corrections of \citet{harvey1997testing}
and \citet{coroneo2020comparing} did not change this. A bootstrap-$t$
version, resampling blocks of rounds when the score differences are
dependent, kept its error near the nominal level in all but the smallest
design, with power similar to that of the test with normal critical
values, slightly lower for the widened interval. In the ES data it removed most of the
tilt towards the less conservative model, although at the 99\% level its
significant results in favour of that model were still the less reliable.
It is the natural replacement when an asymptotic justification suffices. Third, when a guarantee that holds in
finite samples, at all sample sizes and without distributional assumptions
is needed, as when a comparison is monitored over time or its conclusion
carries regulatory weight, use the confidence sequence of
Section~\ref{sec:cs}, tuned to the planned horizon when one is known; in
our simulations the price was two to three times as many rounds as the
Diebold--Mariano test needs for the same power.

For ES, 250-day windows hold too few exceedances for a ranking, and the
Diebold--Mariano test with normal critical values fills the gap in favour
of the less conservative model. Comparative backtests meant to reward a less conservative ES
forecast should state the bound on the loss tail they assume.

The conditions of Theorem~\ref{thm:upper} are needed for the rate, not for
validity. If endpoint distances are heavy-tailed, as when intervals are
scaled by a heavy-tailed volatility, the rate depends on the realized
ranges, and the Diebold--Mariano statistic may itself have a non-Gaussian
limit \citep{frederiksen2026heavy}. The results extend to the weighted
interval score of epidemic forecasting \citep{bracher2021evaluating}, which
is a sum of interval scores and satisfies Condition~\ref{cond:range} with
the summed ranges.

\section*{Data and code availability}
The data sets are public; Section~\ref{sec:empirical} lists their sources,
and the index prices of Section~\ref{sec:es-empirical} are daily closes from
Yahoo Finance, downloaded on 27 October 2026; the downloaded prices are
included with the code.
Code that reproduces every number, table and figure is provided as
Supporting Information and is archived at
\url{https://doi.org/10.5281/zenodo.23770670}
(repository: \url{https://github.com/raphaelrrcoelho/tail-forecast-comparison}).

\section*{Conflict of interest}
The author declares no conflict of interest.

\section*{Funding}
This research received no specific funding.

\appendix

\section{Proofs}
\label{app:proofs}

\subsection{Proof of Lemma~\ref{lem:structure}}

The distance from $y$ to $[l,u]$ is $O(y)=\max\{l-y,0,y-u\}$, and
$\IS_\alpha=(u-l)+(2/\alpha)O$. For real numbers,
$\min_i(x_i-z_i)\le\max_ix_i-\max_iz_i\le\max_i(x_i-z_i)$. Applying this
to $f_t=O_A(y_t)-O_B(y_t)$, with the three terms paired in order, gives
$m_t^-\le f_t\le m_t^+$. The distance to a set is 1-Lipschitz in the
Hausdorff distance between sets, which for intervals equals $H_t$; hence
$|f_t|\le H_t$, and $m_t^\pm$ are bounded in absolute value by $H_t$.
Item~3 holds because both distances vanish. The variance bound follows from
items~2 and~3 and the union bound.

\subsection{Proof of Theorem~\ref{thm:valid}}

Fix $\lambda\ge0$ and a round $t$, write $R=R^-_t$, $D=D_t$, and let
$F_t=\exp\{\lambda(d_t-\Delta_t)-\lambda^2g(\lambda R)D^2\}$ if
$\lambda R<1$ and $F_t=0$ otherwise. We show $\E[F_t\mid\G_t]\le1$. If
$\lambda R\ge1$ there is nothing to prove. If $0<\lambda R<1$, apply the
inequality $\exp\{u\xi-\psi_E(u)\xi^2\}\le1+u\xi$, where
$\psi_E(u)=-\log(1-u)-u$, valid for $\xi\ge-1$ and $u\in[0,1)$
\citep{fan2015exponential}, with $\xi=D/R\ge-1$ and $u=\lambda R$:
\[
 \exp\{\lambda D-\lambda^2g(\lambda R)D^2\}\le1+\lambda D .
\]
If $R=0$, then $D\ge0$ and the same inequality holds with $g(0)=1/2$,
because $e^{x-x^2/2}\le1+x$ for $x\ge0$. Multiplying by
$\exp\{\lambda(c_t-\Delta_t)\}$ and taking conditional expectations, with
$\E[D\mid\G_t]=\Delta_t-c_t$,
\[
 \E[F_t\mid\G_t]\le\{1+\lambda(\Delta_t-c_t)\}e^{-\lambda(\Delta_t-c_t)}\le1 .
\]
Hence $M_t(\lambda)=\prod_{s\le t}F_s$ is a nonnegative supermartingale with
$M_0=1$, and so is the mixture $M_t=\sum_kw_kM_t(\lambda_k)$. Ville's
inequality \citep{ville1939etude} gives
$\Pr(\exists t:M_t\ge2/\delta)\le\delta/2$. Since
$t\Dbar_t=\sum_{s\le t}\Delta_s$, $M_t=\Phi_t(\Dbar_t)$, and
$\Phi_t(\Dbar_t)<2/\delta$ implies $\Dbar_t>L_t$. The upper bound is
symmetric, and a union bound completes the proof.

\subsection{Proof of Theorem~\ref{thm:upper}}
\label{app:upper}

Write $\ell_1=\log(2K/\delta)$, $\ell_2=\log(2/\beta)$, so that
$\ell=\ell_1+\ell_2$, and $Q_t=\sum_{s\le t}D_s^2$. Since
$-R_s^-\le D_s\le R_s^+$, $|D_s|\le\rbar$.

\paragraph{Step 1: a deterministic bound on $\bar d_t-L_t$.}
If $\lambda_k\le1/(2\rbar)$, then $\lambda_kR_s^-\le1/2$ for every $s$, so
the $k$th component of $\Phi_t$ is never set to zero. Every term of
$\Phi_t$ is nonnegative, so for each such $k$,
$\Phi_t(m)\ge K^{-1}\exp\{\lambda_kt(\bar d_t-m)-V_{k,t}\}$, which is at
least $2/\delta$ when $m\le\bar d_t-(\ell_1+V_{k,t})/(\lambda_kt)$. Hence
\[
 L_t\ge\bar d_t-\rho_t,\qquad
 \rho_t=\min_{k:\,\lambda_k\le1/(2\rbar)}\frac{\ell_1+V_{k,t}}{\lambda_kt}.
\]
For these $k$, $g$ being increasing, $V_{k,t}\le g(1/2)\lambda_k^2Q_t$ with
$g(1/2)=4(\log2-1/2)<0.773$. Let $\lambda_*=\{\ell_1/(0.773\,Q_t)\}^{1/2}$
($\lambda_*=\infty$ if $Q_t=0$), and let $\lambda_\sharp$ be the largest
grid point not exceeding $1/(2\rbar)$. Because $\lambda_0\ge1/(4\rbar)$ and
the grid halves down to $\lambda_{K-1}\le1/(\rbar\sqrt t)\le1/(2\rbar)$ (for
$t\ge4$), $\lambda_\sharp\in[1/(4\rbar),1/(2\rbar)]$.

If $\lambda_*\ge\lambda_\sharp$, take $\lambda_k=\lambda_\sharp$:
\[
 t\rho_t\le\frac{\ell_1}{\lambda_\sharp}+0.773\,\lambda_\sharp Q_t
 \le4\rbar\ell_1+0.773\,\lambda_*Q_t
 =4\rbar\ell_1+(0.773\,\ell_1Q_t)^{1/2}.
\]
If $\lambda_*<\lambda_\sharp$, note that $Q_t\le t\rbar^2$ and
$\ell_1\ge\log4>0.773$ (as $\delta\le1/2$) give
$\lambda_*\ge1/(\rbar\sqrt t)\ge\lambda_{K-1}$, so some grid point
satisfies $\lambda_*/2<\lambda_k\le\lambda_*$, and
\[
 t\rho_t\le\frac{2\ell_1}{\lambda_*}+0.773\,\lambda_*Q_t
 =3\,(0.773\,\ell_1Q_t)^{1/2}.
\]
In both cases
\begin{equation}
 t\rho_t\le2.64\,(\ell_1Q_t)^{1/2}+4\rbar\ell_1 .
 \label{eq:rho}
\end{equation}

\paragraph{Step 2: the size of $Q_t$.}
Let $Z_s=D_s^2/\rbar^2\in[0,1]$. Convexity gives $e^z\le1+(e-1)z$ on
$[0,1]$, and $\E[Z_s\mid\G_s]\le\Pr(D_s\ne0\mid\G_s)\le\pi$, so
\[
 \E[e^{Z_s}\mid\G_s]\le1+(e-1)\pi\le e^{(e-1)\pi}.
\]
Thus $\exp\{\sum_{s\le t}Z_s-(e-1)\pi t\}$ has expectation at most one, and
Markov's inequality gives, with probability at least $1-\beta/2$,
\begin{equation}
 Q_t\le\rbar^2\{(e-1)\pi t+\ell_2\}.
 \label{eq:Q}
\end{equation}

\paragraph{Step 3: the deviation of $\bar d_t$.}
The variables $\xi_s=\Delta_s-d_s=\E[D_s\mid\G_s]-D_s$ are martingale
differences with $\xi_s\le R_s^++R_s^-\le2\rbar$ and
$\sum_{s\le t}\E[\xi_s^2\mid\G_s]\le\sum_{s\le t}\E[D_s^2\mid\G_s]\le
t\rbar^2\pi=:v$. Freedman's inequality \citep{freedman1975tail} gives
$\Pr(\sum_{s\le t}\xi_s\ge x)\le\exp[-x^2/\{2(v+2\rbar x)\}]$. With
$x=(2v\ell_2)^{1/2}+4\rbar\ell_2$ the right-hand side is at most
$\beta/2$, so with probability at least $1-\beta/2$,
\begin{equation}
 \bar d_t\ge\Dbar_t-\Bigl(\frac{2\rbar^2\pi\ell_2}{t}\Bigr)^{1/2}
 -\frac{4\rbar\ell_2}{t}.
 \label{eq:freedman}
\end{equation}

\paragraph{Step 4: combination.}
On the intersection of the events in \eqref{eq:Q} and
\eqref{eq:freedman}, which has probability at least $1-\beta$,
\eqref{eq:rho} and $\sqrt{a+b}\le\sqrt a+\sqrt b$,
$\sqrt{\ell_1\ell_2}\le\ell/2$ and $\ell_1,\ell_2\le\ell$ give
\[
 L_t\ge\Dbar_t-\bigl\{\sqrt2+2.64\sqrt{e-1}\bigr\}
 \Bigl(\frac{\rbar^2\pi\ell}{t}\Bigr)^{1/2}
 -\bigl(4+1.32+4\bigr)\frac{\rbar\ell}{t}
 \ge\Delta-4.88\Bigl(\frac{\rbar^2\pi\ell}{t}\Bigr)^{1/2}-9.32\,\frac{\rbar\ell}{t}.
\]
Both subtracted terms are below $\Delta/2$ under \eqref{eq:upper}, which
proves $L_t>0$.

\subsection{Proof of Theorem~\ref{thm:lower}}

Write the score for level $1-\tau$ as
$\rho_{1-\tau}(y-q)=\tau(q-y)+(y-q)_+$. Let $y=q_B$ with probability
$\theta_j$ and $y=q_A-1$ otherwise, under $P_j$. Then
$d=\rho_{1-\tau}(y-q_A)-\rho_{1-\tau}(y-q_B)=-\tau H+H\ind(y=q_B)$, so
$\E_jd=H(\theta_j-\tau)$. Take $\theta_0=\tau$ and $\theta_1=\tau+\Delta/H\le2\tau$.
Under $P_j$, $q_A$ is exceeded with probability $\theta_j$ and $q_B$ never.
The Kullback--Leibler divergence between the two Bernoulli laws is bounded
by their $\chi^2$ divergence,
$\KL(P_1\|P_0)\le(\theta_1-\theta_0)^2/\{\theta_0(1-\theta_0)\}
=\Delta^2/\{H^2\tau(1-\tau)\}$, and for product measures it is $T$ times
this. By the Bretagnolle--Huber inequality
\citep[Lemma~2.6]{tsybakov2009introduction}, the sum of the two error
probabilities of any test is at least $\frac12\exp\{-T\KL(P_1\|P_0)\}$; if
both are at most $\delta$, then $T\KL(P_1\|P_0)\ge\log\{1/(4\delta)\}$. For
the interval version, the lower endpoint is never crossed, so
$\IS_\alpha(I_A;y)-\IS_\alpha(I_B;y)=(2/\alpha)d=d/\tau$.

\subsection{Proof of Proposition~\ref{prop:hoeffding}}

Each component of the mixture is
$\exp\{\lambda t(\bar d_t-m)-\lambda^2tr^2/8\}$ with $\lambda\ge0$. If
$m\ge\bar d_t$, it is at most one, and so is the mixture, which therefore
stays below $2/\delta$. If $m<\bar d_t$, maximizing over $\lambda$ bounds
it by $\exp\{2t(\bar d_t-m)^2/r^2\}$, and so is the mixture, which
reaches $2/\delta$ only if $\bar d_t-m\ge r\sqrt{\log(2/\delta)/(2t)}$.
Hence $L^{\rm H}_t\le\bar d_t-r\sqrt{\log(2/\delta)/(2t)}$. With
$r=2H/\alpha$ the radius is at least $2\Delta$ when
$t\le H^2\log(2/\delta)/(2\alpha^2\Delta^2)$, and then $L^{\rm H}_t>0$
requires $\bar d_t-\Delta>\Delta$, which by Chebyshev's inequality has
probability at most $\sigma^2/(t\Delta^2)$.

\subsection{Proof of Proposition~\ref{prop:dm}}

On the event $N_T$ that $|\varepsilon_t|\le a$ for all $t\le T$, which has
probability $(1-p)^T$ and depends on $(\varepsilon_t)$ only, every round
falls inside both intervals and $d_t=c_t=-2(b-a)\sigma_t$. Then
$\mathrm{DM}_T=-\sqrt T\bar\sigma_T/\hat s_{\sigma,T}$, where
$\hat s_{\sigma,T}$ is the same variance estimator applied to
$(\sigma_t)$. By independence, conditioning on $N_T$ does not change the
law of $(\sigma_t)$. Let $q_T=\Pr(\sqrt T\bar\sigma_T/\hat
s_{\sigma,T}>z_{1-\gamma/2})$, which depends only on the law of
$(\sigma_t)$. The sample autocovariances satisfy
$|\hat\gamma_k|\le\hat\gamma_0$, so the Newey--West estimator with Bartlett
weights and lag $L_T$ satisfies $\hat s_{\sigma,T}^2\le(2L_T+1)\hat\gamma_0$,
and $\hat\gamma_0\le T^{-1}\sum_{t\le T}\sigma_t^2$ is bounded in
probability because $\E\sigma_t^2<\infty$. With $L_T=o(T)$, $\hat
s_{\sigma,T}^2=o_p(T)$, while $\bar\sigma_T\to\E\sigma_1>0$ almost surely,
so $\sqrt T\bar\sigma_T/\hat s_{\sigma,T}\to\infty$ in probability and
$q_T\to1$. If $\sigma_t$ is constant, $\hat s_{\sigma,T}=0$ and the test
rejects by convention.

\subsection{Proof of Proposition~\ref{prop:dm-valid}}

Let $Q$ agree with $P$ on $[-a,a]$ and put the remaining mass $p$ on
$\pm(b+1)$, beyond both intervals. There $d=c+(2/\alpha)(b-a)$ with
$c=-2(b-a)$, so $\E_Qd=2(b-a)(p/\alpha-1)\ge0$, and $A$ is not better
under $Q$; hence $\Pr_Q(\phi\text{ declares }A\text{ better})\le\delta$. Draw
$\varepsilon_t$ from $P$ and replace each value outside $[-a,a]$ by a draw
from the tail of $Q$; this couples $P^T$ and $Q^T$ so that the samples
coincide on the event $N_T$ of no miss, and the total variation distance
between $P^T$ and $Q^T$ is at most $1-(1-p)^T\le Tp$. The bound for $\phi$
follows. On $N_T$, which has probability $(1-p)^T$ under both laws, the
Diebold--Mariano test declares $A$ better by the proof of
Proposition~\ref{prop:dm}.

\subsection{Proofs for Section~\ref{sec:es}}

\paragraph{Lemma~\ref{lem:fz0}.}
Item~1 is \eqref{eq:fz0}. For item~2, $D_t$ is piecewise linear in $y_t$,
equal to zero above both VaRs; below both it has slope $\omega_{B,t}-\omega_{A,t}\le0$
in $y_t$, so it increases as $y_t$ decreases, and its minimum is attained at
$y_t=\min\{v_{A,t},v_{B,t}\}$ or above, where it equals
$-\omega_{B,t}(v_{B,t}-v_{A,t})_+$ or zero. For item~3,
$\partial w/\partial v=1/e<0$, and $\partial w/\partial e=(e-v)/e^2\le0$
for $e\le v$. Moving from $(v_A,e_A)$ to $(v_A,e_B)$ keeps $e\le e_A\le v_A$
and does not decrease $w$; moving on to $(v_B,e_B)$ increases $w$, strictly
if $v_A>v_B$.

\paragraph{Proposition~\ref{prop:es-one-sided}.}
On a day in $\mathcal L$, $D_t$ is nonincreasing in $y_t$ below
$\min\{v_{A,t},v_{B,t}\}$ by Lemma~\ref{lem:fz0}, so $\tilde d_t\le d_t$
and $\E[\tilde d_t\mid\G_t]\le\Delta_t$. The process of the proof of
Theorem~\ref{thm:valid}, built from $\tilde d_t$ and with factor one on days
outside $\mathcal L$, is a nonnegative supermartingale because $\mathcal L$
is predictable, and the argument of that theorem bounds the average of
$\E[\tilde d_t\mid\G_t]$, hence of $\Delta_t$, from below. The statement
for the complement is symmetric.

\paragraph{Proposition~\ref{prop:es-impossible}.}
For $\varepsilon\in(0,1)$ and $M>0$ let $P_{\varepsilon,M}$ be the
mixture of $P_1$, with weight $1-\varepsilon$, and a point mass at $-M$,
with weight $\varepsilon$. By
Lemma~\ref{lem:fz0}, $d$ evaluated at $y=-M$ grows linearly in $M$ with
slope $\omega_A-\omega_B>0$, so for $M$ large enough $\E_{P_{\varepsilon,M}}d\ge0$
and $P_{\varepsilon,M}$ belongs to the hypothesis. The total variation
distance between the $T$-fold products of $P_1$ and $P_{\varepsilon,M}$ is
at most $T\varepsilon$. Hence $P_1(\phi\text{ rejects})\le\delta+T\varepsilon$
for every $\varepsilon>0$.

\section{The tuned sequence and its power}
\label{app:tuned}

The tuned sequence estimates $v$, the mean of $D_t^2$, and the 90th
percentiles $r^\mp$ of $R_t^\mp$ on the tuning segment. For each side it
sets $\lambda_*=\{2\log(4/\delta)/(nv)\}^{1/2}$, capped at $1/(2r^\mp)$,
with $n$ the number of remaining rounds, and uses the grid
$\lambda_*2^j$, $j=-10,\ldots,9$, with weight $1/2$ on $j=0$ and $1/38$ on
each other point. For a component with weight $w$, the lower bound is at
least $\bar d_t-\{\log(2/(\delta w))+V_{k,t}\}/(\lambda_kt)$, which is
smallest near this $\lambda_*$ when $V_{k,t}\approx\lambda_k^2nv/2$.
Validity follows from Theorem~\ref{thm:valid}, since the weights are fixed
before the remaining rounds are observed; the target is the average of
$\Delta_t$ over those rounds.

In practice:
\begin{enumerate}
\item compute $d_t$, $c_t$, $a_t$ and $b_t$ for every round from the
 forecasts and the outcomes (Lemma~\ref{lem:structure} for intervals,
 \eqref{eq:quantile} for quantiles, Lemma~\ref{lem:fz0} for (VaR, ES));
\item on the tuning segment, set $v$ to the mean of $D_t^2$ and $r^\mp$
 to the 90th percentiles of $R_t^\mp$, and form the grid and weights above
 for each side;
\item on the remaining rounds, maintain $\sum_s d_s$ and, for each $k$,
 $\sum_s g(\lambda_kR^\mp_s)D_s^2$, and find $L_t$ and $U_t$ by bisection
 on $m$ in $\Phi_t(m)=2/\delta$;
\item declare $B$ better once $L_t>0$, and $A$ better once $U_t<0$.
\end{enumerate}
The code accompanying the paper implements these steps.

Table~\ref{tab:power} compares its power with that of the Diebold--Mariano
test and its bootstrap-$t$ version in the Gaussian design of
Section~\ref{sec:sim}.

\begin{table}[h]
\centering\small
\caption{Power (\%) to declare the worse forecaster $A$ worse in the
Gaussian design of Figure~\ref{fig:detection}, at horizons $T_{80}$,
$2T_{80}$ and (for EB) $4T_{80}$, where $T_{80}$ gives the known-variance test 80\% power. DM:
two-sided 5\% Diebold--Mariano test; boot: its bootstrap-$t$ version;
EB: the tuned sequence, tuned on the first 10\% of the horizon. Neither
DM nor boot declared $A$ better in any replication. 200 replications.}
\label{tab:power}
\setlength{\tabcolsep}{4pt}\begin{tabular}{rrrrrrrrrrrrrrr}
\toprule
 & \multicolumn{7}{c}{$A$ narrowed} & \multicolumn{7}{c}{$A$ widened}\\
\cmidrule(lr){2-8}\cmidrule(lr){9-15}
 & \multicolumn{3}{c}{$T=T_{80}$} & \multicolumn{3}{c}{$T=2T_{80}$} & $4T_{80}$ & \multicolumn{3}{c}{$T=T_{80}$} & \multicolumn{3}{c}{$T=2T_{80}$} & $4T_{80}$\\
$\alpha$ & DM & boot & EB & DM & boot & EB & EB & DM & boot & EB & DM & boot & EB & EB\\
\midrule
0.2 & 79.0 & 83.0 & 12.5 & 98.5 & 99.0 & 56.0 & 98.0 & 79.0 & 70.5 & 16.0 & 97.5 & 96.0 & 48.5 & 91.5\\
0.1 & 79.0 & 83.5 & 24.0 & 98.5 & 99.0 & 76.0 & 99.5 & 74.0 & 68.5 & 18.5 & 95.5 & 94.5 & 46.5 & 94.5\\
0.05 & 81.5 & 83.0 & 28.5 & 99.0 & 99.5 & 75.5 & 99.0 & 83.0 & 76.5 & 17.5 & 96.0 & 95.5 & 64.0 & 98.5\\
0.02 & 79.5 & 82.5 & 34.5 & 97.0 & 98.5 & 77.5 & 100.0 & 80.0 & 75.0 & 20.5 & 97.0 & 95.0 & 63.0 & 98.5\\
0.01 & 81.0 & 84.5 & 37.5 & 99.5 & 100.0 & 82.5 & 100.0 & 80.0 & 73.0 & 28.5 & 96.5 & 94.5 & 58.5 & 98.0\\
0.005 & 83.5 & 87.5 & 33.0 & 99.0 & 99.5 & 88.0 & 99.0 & 75.5 & 69.5 & 28.0 & 95.0 & 95.0 & 66.5 & 96.5\\
\bottomrule
\end{tabular}

\end{table}

\bibliographystyle{abbrvnat}
\bibliography{refs}

\end{document}